\documentclass[onecolumn,secnumarabic,amsmath,amssymb,balancelastpage,nofootinbib]{article}

\usepackage[e]{esvect}
\usepackage{color}         
\usepackage{graphics}      
\usepackage{graphicx}      
\usepackage{epsf}          
\usepackage{bm}            

\usepackage{cite}
\usepackage{amssymb}
\usepackage{amsmath, esint}
\usepackage{mathrsfs}
\usepackage{framed}
\usepackage{bigints} 
\usepackage{enumitem}
\usepackage{pifont}
\usepackage[capbesideposition={left,center},facing=yes,capbesidewidth=8cm,capbesidesep=quad]{floatrow} 
\usepackage{setspace} 

\usepackage[none]{hyphenat} 

\usepackage[colorlinks=true]{hyperref}  

\usepackage{feynmf}

\definecolor{darkred}{rgb}{0.6,0,0}
\definecolor{darkgreen}{rgb}{0,0.5,0}
\definecolor{darkblue}{rgb}{0,0,0.6}
\hypersetup{ colorlinks,
linkcolor=darkblue,
filecolor=darkgreen,
urlcolor=darkgreen,
citecolor=darkred }

\newcommand{\del}[0]{\ensuremath{\vec{\nabla}}}

\begin{document}

\sloppy 

\bibliographystyle{nar}

\newlength{\bibitemsep}\setlength{\bibitemsep}{.2\baselineskip plus .05\baselineskip minus .05\baselineskip}
\newlength{\bibparskip}\setlength{\bibparskip}{0pt}
\let\oldthebibliography\thebibliography
\renewcommand\thebibliography[1]{%
  \oldthebibliography{#1}%
  \setlength{\parskip}{\bibitemsep}%
  \setlength{\itemsep}{\bibparskip}%
}


\title{\vspace*{-35 pt}\huge{The Disappearance and Reappearance of Potential Energy in Classical and Quantum Electrodynamics}}
\author{Charles T. Sebens\\Division of the Humanities and Social Sciences\\ California Institute of Technology}
\date{September 30, 2022\\arXiv v.3}

\maketitle
\vspace*{-20 pt}
\begin{abstract}
In electrostatics, we can use either potential energy or field energy to ensure conservation of energy. In electrodynamics, the former option is unavailable. To ensure conservation of energy, we must attribute energy to the electromagnetic field and, in particular, to electromagnetic radiation. If we adopt the standard energy density for the electromagnetic field, then potential energy seems to disappear. However, a closer look at electrodynamics shows that this conclusion actually depends on the kind of matter being considered. Although we cannot get by without attributing energy to the electromagnetic field, matter may still have electromagnetic potential energy. Indeed, if we take the matter to be represented by the Dirac field (in a classical precursor to quantum electrodynamics), then it will possess potential energy (as can be seen by examining the symmetric energy-momentum tensor of the Dirac field). Thus, potential energy reappears. Upon field quantization, the potential energy of the Dirac field becomes an interaction term in the Hamiltonian operator of quantum electrodynamics.
\end{abstract}

\renewcommand{\baselinestretch}{1.0}\normalsize
\tableofcontents
\renewcommand{\baselinestretch}{1.3}\normalsize

\section{Introduction}\label{introsec}

Are potential energies part of fundamental physics?  This is a difficult question with twists and turns.  Confining our attention to electrostatics, we can freely choose between attributing potential energy to matter or, instead, attributing that energy to the electric field (section \ref{ESsection}).  We can use either potential energy or field energy to balance the work done on matter by electric forces and ensure conservation of energy.  In full electrodynamics, we cannot use potential energy alone to balance the work done by electromagnetic forces.  If energy is to be conserved, we must attribute energy to the electromagnetic field and, in particular, to electromagnetic radiation (section \ref{EMsection}).  The standard way of doing so uses field energy alone to balance the work done by electromagnetic forces.  This is all familiar ground.\footnote{See \cite[ch.\ 27]{feynman2}; \cite[ch.\ 5]{lange}; \cite[sec.\ 2.4.4]{griffiths}.}

Pausing here, potential energy seems to have disappeared from fundamental physics.  But, a closer look at electrodynamics shows that this conclusion actually depends on the kind of matter being considered.  Although we cannot get by without attributing energy to the electromagnetic field, matter may still have electromagnetic potential energy.  Indeed, if we model matter using the classical Dirac field, its energy and momentum will depend on the state of the electromagnetic field---as can be seen by examining the symmetric energy-momentum tensor for the Dirac field (section \ref{MDsection}).  Thus, potential energy reappears.

There is some room for debate as to which terms in the Dirac field's energy density give the field's potential energy density.  This issue can be better understood by considering the Aharonov-Bohm effect (appendix \ref{KPED}).

The existence of potential energy in the classical theory of interacting Dirac and electromagnetic fields yields an interaction term in the Hamiltonian operator when we quantize these fields and move to quantum electrodynamics---the quantum field theory of electrons, positrons, and photons (section \ref{QFTsection}).  To streamline the transition from classical physics to quantum field theory, our focus throughout will be on continuous distributions of mass and charge (as opposed to point particles).

For the most part, the various pieces of the account presented in this article can be found within physics textbooks.  However, the pieces are scattered and I have not seen them assembled to address the ontological status of potential energy.  In particular, I have not seen authors confront the mystery as to how potential energy could reemerge for the Dirac field after it had apparently been excised in classical electromagnetism.

There are many reasons that could be given for asking what becomes of potential energy as we move from electrostatics to more fundamental physics.  Here are a few:  First, it is generally valuable to get clear on the relationships between different theories in physics.  It is important to know what is retained and what is jettisoned in the move from one theory to another.  New theories are often presented as radical departures from old theories, but I tend to think that novelties are often overstated and that continuities are often underappreciated.  Here, I argue that potential energy survives in quantum electrodynamics (even though it is not generally described as such).  Second, studying the nature of potential energy and field energy can help us articulate and evaluate particular proposals for the laws and ontology of classical and quantum electrodynamics.  For example, in classical electrodynamics the fact that the electromagnetic field must carry energy to achieve conservation of energy has been used to argue that the electromagnetic field is a real thing (just like charged matter) \cite{lange}; \cite[sec.\ 4.2]{lazarovici2018}.  Third, given that the electromagnetic field is a real thing possessing energy and momentum, one might wonder how far the similarities between matter and field go.  Like matter, the electromagnetic field possesses mass, has a velocity at each point in space, and experiences forces (equal and opposite the Lorentz forces exerted on matter \cite{forcesonfields}).  Here I present a disanalogy in the context of a classical or quantum theory of electromagnetic and Dirac fields: the Dirac field has potential energy but the electromagnetic field does not.  Fourth, a better understanding of potential energy and field energy in classical and quantum electrodynamics could be useful for addressing puzzles regarding self-interaction, self-energy, and self-fields in quantum electrodynamics \cite{feynman1965, barut1988, barut1990, blumjoas, blum2017, selfrepulsion, fundamentalityoffields}.  In section \ref{MDsection}, we will briefly discuss the potential energy of an electron interacting with its own electromagnetic field.

\section{Potential Energy}\label{potsec}

What distinguishes potential energy from other forms of energy?  When the concept of potential energy was first introduced, potential energy was contrasted with actual energy \cite{hecht2003, roche2003}.  That contrast fits poorly with contemporary usage, where potential energy is regarded as real energy.  If we want the total amount of actual energy to be conserved in a theory that includes potential energy, we must treat potential energy as actual energy.\footnote{Sir John Herschel and Ernst Mach regarded potential energy as unreal energy and reacted to the above problem by downgrading conservation of energy, with Mach describing it as a mere ``housekeeping principle'' \cite[pg.\ 191]{roche2003}.}

Instead of contrasting potential energy with actual energy, one might contrast potential energy with kinetic energy. If we can identify the kinetic energy in a given theory, then we might classify all other energy as potential \cite[sec.\ 7]{roche2003}.  Potential energy would then be a catch-all with no feature to distinguish itself other than being non-kinetic.  On this view, we would be forced to categorize all forms of energy as either kinetic or potential, including the energy of the electromagnetic field in classical electrodynamics and the $mc^2$ rest mass energy in relativistic particle mechanics.  Perhaps all energies can be sorted into these two boxes, but it is limiting to rule out the possibility of other forms of energy.

Another idea is to link potential energy to forces.  Suppose that we have a conservative force where the work done by that force on a collection of bodies depends only on the initial and final arrangement of those bodies in space (as would be the case for gravitational forces in Newtonian gravity or electric forces in electrostatics).  Then, we can define the total potential energy of the collection of bodies in a particular configuration to be the work that must be done against the force in question to move the bodies into that configuration from some stipulated reference configuration (such as the bodies being infinitely far apart).\footnote{Lange \cite[ch.\ 5]{lange} discusses this idea and its relation to field energy.}  Looking ahead to where we are going, there are two problems with this idea.  First, in sections \ref{EMsection} and \ref{MDsection} we will be considering non-conservative forces and I do not want to rule out the idea that matter might possess some kind of potential energy automatically.  Second, this understanding of potential energy fails to differentiate between potential energy and field energy.  In electrostatics, the work required to bring a set of point charges in from infinity to a particular configuration---or to assemble a particular density of charge---is straightforward to calculate.  However, as we will see in the next section, there remains room for debate as to whether such processes should be understood as increasing the potential energy of charged matter or as increasing the energy stored in the electric field.

For our purposes here, let us take the distinguishing feature of potential energy to be the fact that it is \emph{extrinsic}.\footnote{By contrast, kinetic energy appears to be intrinsic (though that is not to say that it appears to be the only kind of energy that is intrinsic).  The kinetic energy of a point particle in a particular frame is fixed by its mass and velocity in that frame.  Thus, whether a particle's kinetic energy is intrinsic turns on whether its mass and velocity are intrinsic properties.  There are reasons why one might question whether the velocity of a particle is an intrinsic property of that particle.  First, there has been philosophical debate as to whether a body's velocity is an intrinsic property of that body \emph{at a particular instant in time} \cite{arntzenius2000, smith2003, lange2005}.  Still, velocity might be an intrinsic property of a body even if it requires considering a time interval and thus is not an intrinsic property of that body at a moment.  Second, one could argue that the velocity of a particle is not intrinsic because it depends on the relation of that body to the background spacetime in which it is moving.  For our purposes here, let us put this concern aside and count velocity as intrinsic-enough to support a useful distinction between a particle's intrinsic kinetic energy and any extrinsic potential energy it may have.\label{KEfootnote}}  The potential energy of an entity is not solely dependent on the physical state of that entity alone.  For example, in a Newtonian theory of gravity one might attribute potential energy to a ball that depends on its distance from the earth's center.  Alternatively, we might say that the ball's potential energy depends on the strength of the gravitational field at the ball's location.  Either way, the potential energy of the ball depends on something other than the ball---the earth below the ball or the gravitational field within the ball.  In the following sections, we will see how this understanding of potential energy as extrinsic energy can be applied in different theoretical contexts.  (If you do not agree that potential energy is extrinsic energy, you may read this paper as an exploration of intrinsic and extrinsic energy---asking whether there is a separate intrinsic energy density for the electromagnetic field and another for charged matter.\footnote{In section \ref{MDsection}, will see that if charged matter is modeled by the classical Dirac field then its energy density depends on the electromagnetic field and is thus not entirely intrinsic.  We can identify a potential energy density that captures the dependence on the state of the electromagnetic field.  Although it is not apparent from just studying interactions between the Dirac and electromagnetic fields, the energy density of the electromagnetic field is arguably also not entirely intrinsic because, within general relativity, it depends on the gravitational field (or, you might say, on spacetime structure) \cite{lehmkuhl2011}.  That context is beyond the scope of this article, but I would be inclined to say that in general relativity the electromagnetic field has a gravitational potential energy density.})

In addition to potential energy, you sometimes see authors denoting a particular kind of energy as an ``interaction energy.''  If potential energy is energy possessed by one entity that depends on the state of another, interaction energy might naturally be interpreted as energy possessed jointly by two entities.  For example, we might reinterpret the above potential energy of a ball above the earth as an interaction energy possessed by the ball and the gravitational field together.  The distinction between potential energy and interaction energy is subtle and perhaps too metaphysical for some readers.  Certainly, it will often be difficult to settle whether a particular energy should be interpreted as a potential energy or an interaction energy.  However, there can be reasons for preferring one interpretation over another (as we will see in section \ref{MDsection} when we study the energy of the Dirac field).

With this groundwork in place, let us analyze the status of potential energy in different physical theories.

\section{Electrostatics}\label{ESsection}

In electrostatics, we can achieve conservation of energy by attributing an energy density to the electric field, a potential energy density to matter, or an interaction energy density to matter and field together.  These densities can also be combined.  In this section, we will review the wide array of options.  Gaussian cgs units will be used throughout.

In electrostatics, the force from a point charge $Q$ on a point charge $q$ is
\begin{equation}
\vec{F}=\frac{Qq}{r^2} \hat{u}
\ ,
\label{eforceexample}
\end{equation}
where $\hat{u}$ is a unit vector pointing from $Q$ to $q$ and $r$ is the distance between $Q$ and $q$.  We can separate this equation for the force on $q$ into an equation for the electric field $\vec{E}$ sourced by $Q$ at the location of $q$,
\begin{equation}
\vec{E}=\frac{Q}{r^2} \hat{u}
\ ,
\label{efieldpointcharge}
\end{equation}
and an equation for the force on $q$ from the electric field,
\begin{equation}
\vec{F}=q \vec{E}
\ .
\label{eforcepointcharge}
\end{equation}

Generalizing \eqref{efieldpointcharge} and \eqref{eforcepointcharge} to a distribution of charge $\rho^q$, we can write the laws of electrostatics as
\begin{align}
\del\cdot\vec{E}&=4\pi \rho^q
\label{egauss}
\\
\del\times\vec{E}&=0
\label{efaraday}
\\
\vec{f} &= \rho^q \vec{E}
\ ,
\label{eforcedensity}
\end{align}
where $\vec{f}$ is the density of electric force.  The first two equations describe the way that charge acts a source for the electric field and the last equation gives the density of electric force on matter.  Because the electric field has no curl, it can be written as the gradient of a scalar field $\phi$ (the scalar potential),
\begin{equation}
\vec{E}=-\del \phi
\ ,
\label{epotential}
\end{equation}
where the minus sign is conventional.  Equations \eqref{egauss}--\eqref{eforcedensity} give a general theory of electric interactions that serves as a good approximation to electromagnetism in static situations and also in dynamic situations where magnetic fields can be ignored.  Thus, let us think of electrostatics as a theory of dynamics, not just statics (as the name suggests).

The flow of charge can be described by a velocity field $\vec{v}^{\,q}$ such that the current density\footnote{Noting that neutral wires (with $\rho^q=0$) can carry non-zero currents, Griffiths \cite[pg.\ 357]{griffiths} suggests in a footnote that we introduce separate charge densities and velocity fields for positive and negative charges, writing the current as $\vec{J} = \rho^q_+ \vec{v}^{\,q}_+ + \rho^q_- \vec{v}^{\,q}_-$.  This is a good idea, but to keep things simple I have left this complication out of the above equations.} is given by $\vec{J}=\rho^q \vec{v}^{\,q}$ and local conservation of charge is expressed by the continuity equation
\begin{equation}
\frac{\partial \rho^q}{\partial t} = - \del \cdot (\rho^q \vec{v}^{\,q})
\ .
\label{qcontinuity}
\end{equation}
The rate at which work is done on matter by electric forces per unit time and unit volume (the power density) is the force density times the velocity field, $\vec{f} \cdot \vec{v}^{\,q}$.  Integrating this power density over all of space yields the change in the energy of matter per unit time (setting aside electric potential energy, if there is such a thing),
\begin{align}
\int \vec{f} \cdot \vec{v}^{\,q}\ d^3 x&=\int \rho^q \vec{E} \cdot \vec{v}^{\,q} \ d^3 x
\nonumber
\\
&=\int \vec{E} \cdot \vec{J} \ d^3 x
\ .
\end{align}
Using \eqref{egauss}, \eqref{epotential}, and \eqref{qcontinuity} and integrating by parts (dropping boundary terms), we can derive two distinct equations for the global conservation of energy,
\begin{align}
\int \vec{f} \cdot \vec{v}^{\,q}\ d^3 x\ +\ &\frac{d}{d t} \int \frac{E^2}{8 \pi}\, d^3 x=0
\label{econservation1}
\\
\int \vec{f} \cdot \vec{v}^{\,q}\ d^3 x\ +\ &\frac{d}{d t} \int \frac{1}{2}\rho^q\phi\, d^3 x=0
\ .
\label{econservation2}
\end{align}
The first equation \eqref{econservation1} can be interpreted as assigning the electric field an energy density that is proportional to the square of the field strength,
\begin{equation}
\rho^{\mathcal{E}}=\frac{E^2}{8 \pi}
\ .
\label{efieldenergydensity}
\end{equation}
The rate at which the electric field's energy changes balances the rate at which energy is transferred from the field to matter so that total energy is conserved.  The second equation \eqref{econservation2} can be interpreted as assigning to matter a potential energy density of
\begin{equation}
\rho^{\mathcal{E}}=\frac{1}{2}\rho^q\phi
\ ,
\label{epotentialenergydensity}
\end{equation}
that is dependent on the state of the electric field (as specified by the electric potential).\footnote{These two energy densities for electrostatics are discussed in, e.g., \cite[sec.\ 8.5]{feynman2}; \cite{peters1981}; \cite[sec.\ 2.4]{griffiths}; \cite[sec.\ 3.6]{zangwill2012}.}

Alternatively, one could say that the potential energy of matter at a point depends on the distribution of charge everywhere else, rewriting \eqref{epotentialenergydensity}, for a particular choice of $\phi$, as
\begin{equation}
\rho^{\mathcal{E}}(\vec{x})=\frac{1}{2}\int{\frac{\rho^q(\vec{x}) \rho^q(\vec{x}')}{|\vec{x}-\vec{x}'|}\, d^3 x'}
\ ,
\label{epotentialenergydensity2}
\end{equation}
where here it is convenient to explicitly denote the dependence of the charge and energy densities on the location $\vec{x}$.  The above expression can be thought of as an extension of the familiar $\frac{Qq}{r}$ potential energy for a pair of point charges $q$ and $Q$ to a distribution of charge $\rho^q$.  Using \eqref{epotentialenergydensity2} and understanding the potential energy at a point to be dependent on the density of charge elsewhere allows us to avoid attributing energy to the electric field and even avoid attributing an energy to matter that depends explicitly on the state of the electric field.  Given that the electric field can be eliminated from the energy density, one might attempt to eliminate it entirely---arguing that the electric field is just a useful fiction, not a real thing.  By contrast, in full electromagnetism we will see that the electromagnetic field cannot be eliminated from the energy density.  We have to attribute energy to the electromagnetic field.  This difference between electrostatics and electromagnetism provides an argument for the reality of the electromagnetic field: to achieve conservation of energy, we need a real electromagnetic field that possesses energy.  There is more to be said about the merits of such an argument,\footnote{See \cite{lange}; \cite[sec.\ 4.2]{lazarovici2018} for discussion of this argument and a similar argument from conservation of momentum.} but for our purposes here we will take the conclusion to be correct and treat both matter and electromagnetic field as real things in the sections that follow.

There are a couple other ways that one might interpret the second equation for conservation of energy \eqref{econservation2}.  First, instead of viewing \eqref{epotentialenergydensity} as an energy density possessed by matter and dependent on the electric field, it can be interpreted as an interaction energy density possessed by matter and field together.  Second, using \eqref{egauss} and \eqref{epotential} one can rewrite \eqref{epotentialenergydensity} solely in terms of field variables as $-\frac{\phi\nabla^2\phi}{8 \pi}$ and interpret it as an alternative density of energy for the electric field---different from \eqref{efieldenergydensity}.

Although \eqref{efieldenergydensity} and \eqref{epotentialenergydensity} give different distributions of energy in space, they agree on the total electric energy---provided we choose a potential $\phi$ that goes to zero at infinity, as in \eqref{epotentialenergydensity2}.  If we choose a potential that differs by the addition of a global constant (a gauge transformation), the two expressions will disagree on the total electric energy but will agree on the rate at which that total electric energy changes and will agree that energy is conserved.  The energy density in \eqref{epotentialenergydensity} is gauge-dependent whereas the energy density in \eqref{efieldenergydensity} is not.  This is a virtue of \eqref{efieldenergydensity} and a reason to prefer field energy to potential energy.

One can form additional alternative energy densities that would ensure conservation of total energy by taking linear combinations of \eqref{efieldenergydensity} and \eqref{epotentialenergydensity},
\begin{equation}
\rho^{\mathcal{E}}=\alpha\frac{E^2}{8 \pi} + (1-\alpha)\frac{1}{2}\rho^q\phi
\ ,
\label{emixedenergydensity}
\end{equation}
for any $\alpha$.  With $\alpha$ not equal to zero or one, this can be understood as assigning energy to the electric field and also potential energy to matter.  That might seem like an unnatural move, but the idea of having both field energy and potential energy will be proved prescient in section \ref{MDsection}.

\begin{figure}[htb]
\center{\includegraphics[width=11 cm]{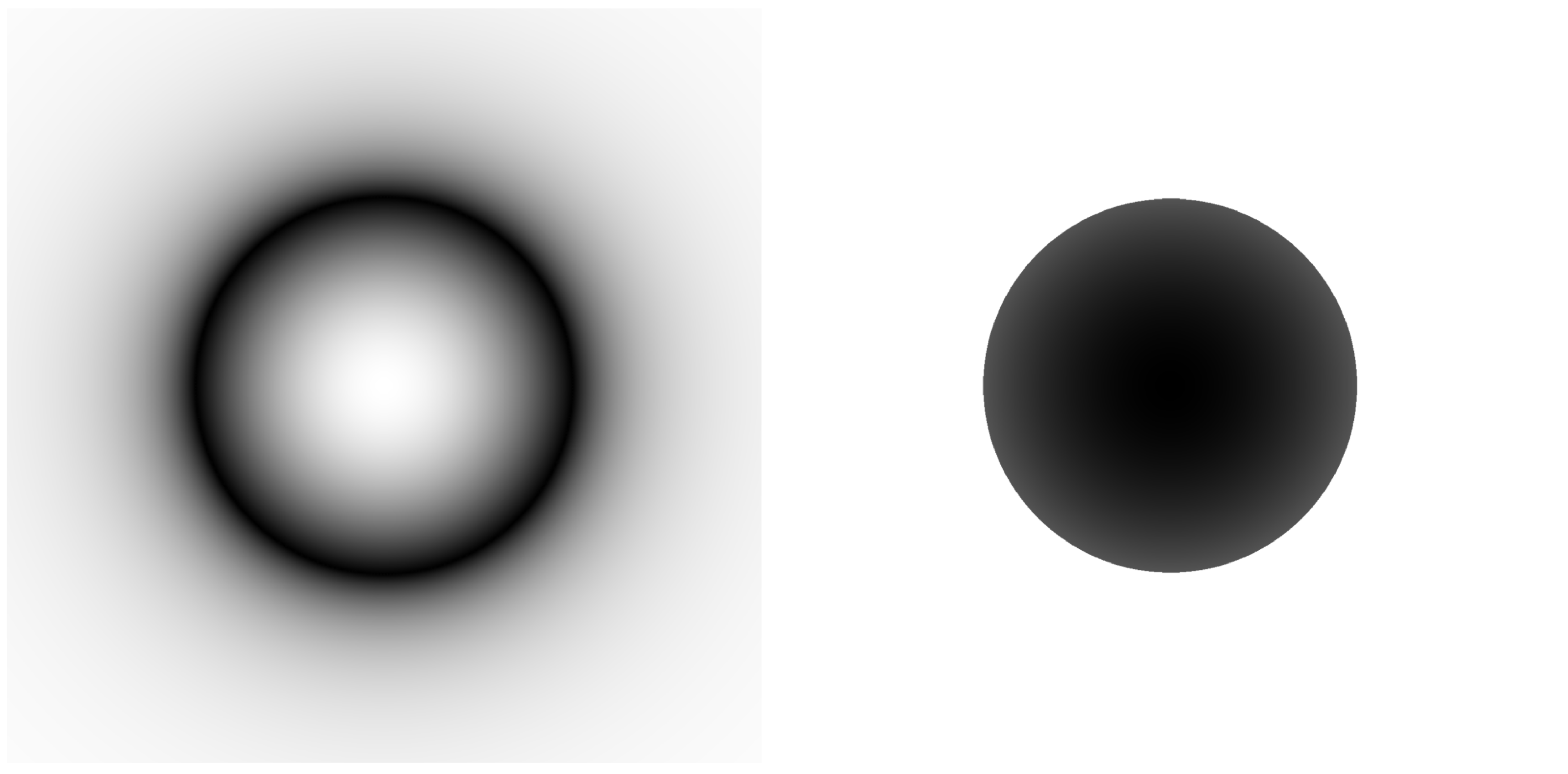}}
\caption{For a sphere of uniform charge density, the electric energy density is plotted on the left using the expression for field energy density in \eqref{efieldenergydensity} and on the right using the expression for potential energy density in \eqref{epotentialenergydensity}.}
\label{energyfigure}
\end{figure}

To compare the energy densities in \eqref{efieldenergydensity} and \eqref{epotentialenergydensity}, consider a sphere with charge distributed uniformly throughout the volume of the sphere (so that the charge density is constant within the bounds of the sphere and zero outside the sphere).  According to \eqref{efieldenergydensity}, the energy density is greatest in magnitude on the sphere's surface (where the electric field is strongest).  The energy density weakens as one moves in towards the sphere's center or out into the empty space surrounding it.  Working through the math, two-thirds of the sphere's electric energy resides in the empty space surrounding it.  By contrast, the energy density in \eqref{epotentialenergydensity} (using an expression for $\phi$ that approaches zero at infinity) is greatest in magnitude at the sphere's center and decreases as one approaches the sphere's surface, dropping to zero beyond that surface (where $\rho^q=0$).  The total electric energy is the same either way, but the spatial distribution of energy is different (figure \ref{energyfigure}).

Here is how Griffiths describes our predicament in his textbook (inserting the relevant equations in the notation used here),
\begin{quote}
``Where \emph{is} the energy then?  Is it stored in the field, as [$\mathcal{E}=\int \frac{E^2}{8 \pi}\, d^3 x$] seems to suggest, or is it stored in the charge, as [$\mathcal{E}=\int \frac{1}{2}\rho^q\phi\, d^3 x$] implies?  At the present stage [in the context of electrostatics] this is simply an unanswerable question: I can tell you what the total energy is, and I can provide you with several different ways to compute it, but it is impertinent to worry about \emph{where} the energy is located.  In the context of radiation theory \dots\ it is useful (and in general relativity it is \emph{essential}) to regard the energy as stored in the field, with a density [$\frac{E^2}{8 \pi}$] \dots'' \cite[pg.\ 97]{griffiths}
\end{quote}
Within electrostatics, we have seen that there are many options for understanding the nature of electric energy.  In electrodynamics, we will see that we must regard some energy as stored in the electromagnetic field if we are to achieve conservation of energy.  However, Griffiths may be right that it is merely ``useful'' to attribute an energy of $\frac{E^2}{8 \pi}$ to the electric field in this context, as there are multiple ways you might assign an energy density to the electromagnetic field---with $\frac{E^2}{8 \pi}$ featuring as the contribution from the electric field in what is arguably the most appealing way of doing so.  General relativity may be taken to settle the matter, for reasons that will be mentioned in the next section.  Reading the quote from Griffiths, it looks like potential energy is an option in electrostatics that disappears when we move to electrodynamics (at least, the option disappears when we incorporate some hindsight from general relativity to settle any lingering indeterminacy about the distribution of energy).  Later, we will see potential energy reappear.

Before proceeding to electrodynamics, let us briefly compare the situation in electrostatics to Newtonian gravity.  The laws of Newtonian gravity take the same form as the laws of electrostatics \eqref{egauss}--\eqref{eforcedensity}.  For a mass distribution $\rho^m$ and gravitational field $\vec{g}$ (expressible in terms of a gravitational potential $\phi_g$ as $\vec{g}=-\del \phi_g$), the laws can be written as $\del\cdot\vec{g}= - 4 \pi G \rho^m$, $\del\times\vec{g}=0$, and $\vec{f} = \rho^m \vec{g}$ (see \cite[sec.\ 7.6]{doughty1990}). In analogy to \eqref{efieldenergydensity}, \eqref{epotentialenergydensity}, and \eqref{emixedenergydensity}, one can achieve conservation of energy by introducing a (negative) gravitational field energy density of $\frac{-g^2}{8 \pi G}$, a potential energy density of $\frac{1}{2}\rho^m\phi_g$, or by taking a linear combination of the two as in \eqref{emixedenergydensity}.\footnote{For discussion of these candidate energy densities for Newtonian gravity, see \cite[sec.\ 1.2]{ohanian1976}; \cite{peters1981}; \cite[box 13.4]{thorneblandford}; \cite{bengtsson2021}; \cite[sec.\ 3.1]{gravitationalfield}.  One might hope that general relativity would give us the benefit of hindsight, telling us which of the available energy densities accurately describes the distribution of gravitational energy in the Newtonian limit.  Unfortunately, the status of gravitational energy in general relativity is a thorny issue that remains unsettled.  There are multiple candidate energy-momentum tensors that can be constructed and existing proposals disagree on the energy density in the Newtonian limit \cite{peters1981, lyndenbell1985}; \cite[sec.\ 5.1]{gravitationalfield}.}  Figure \ref{energyfigure} could be repurposed to depict possible densities of gravitational energy around a planet.

\section{Classical Electrodynamics}\label{EMsection}

In the full theory of electromagnetism, classical electrodynamics, we cannot preserve conservation of energy by simply assigning potential energy to matter or interaction energy to matter and field.  We must attribute energy to the electromagnetic field itself.  In this section, we will begin by reviewing a simple example to illustrate the previous point and then discuss two ways of deriving the field energy density: first from Maxwell's equations and the Lorentz force law (Poynting's theorem) and second from the Lagrangian density (where the energy density can be derived as part of the symmetric energy-momentum tensor for the electromagnetic field).

There is no way to achieve conservation of energy by introducing an electromagnetic energy that depends only on the properties of matter and field at the locations where matter is present, as in \eqref{epotentialenergydensity}---or by introducing an electromagnetic energy that depends only on the arrangement of matter, as in \eqref{epotentialenergydensity2}.  To see why, imagine inputting energy to accelerate a charged body from rest and then taking out energy to decelerate it back to rest at a new location.  The inputted energy will have to be larger than the outputted energy because electromagnetic radiation is produced during the periods of acceleration and deceleration.  By attributing energy to the electromagnetic waves that are emitted, we can ensure that energy is conserved in this process.  However, there is no way to achieve energy conservation by instead attributing a potential energy to the charged body or an interaction energy to the body and the field where it is located.  After the motion, the state of the body and the electromagnetic field inside it are unchanged (just shifted over).  Looking locally, there is no evidence of the past motion.  There are ways of achieving conservation of energy without attributing energy to the electromagnetic field if one is willing to let the potential energy at one moment depend on the past (and perhaps the future), but I will not explore that kind of radical approach here.\footnote{This perspective on potential energy pairs well with approaches to electromagnetism where one attempts to remove the electromagnetic field and have particles interact directly with one another across spatiotemporal gaps (such as Wheeler-Feynman electrodynamics).  See \cite{wheelerfeynman1949}; \cite[pg.\ 80]{ohanian1976}; \cite[ch.\ 5]{lange};  \cite[sec.\ 4.2]{lazarovici2018}.}

The laws of electromagnetism can be expressed in a similar form to the laws of electrostatics \eqref{egauss}--\eqref{eforcedensity} given earlier,
\begin{align}
\vec{\nabla}\cdot\vec{E}&=4\pi \rho^q
\label{gauss}
\\
\vec{\nabla}\times\vec{E}&=-\frac{1}{c}\frac{\partial \vec{B}}{\partial t}
\label{faraday}
\\
\vec{\nabla}\cdot\vec{B}&=0
\\
\vec{\nabla}\times\vec{B}&=\frac{4\pi}{c}\vec{J}+\frac{1}{c}\frac{\partial \vec{E}}{\partial t}
\label{ampere}
\\
\vec{f} &= \rho^q \left(\vec{E} + \frac{1}{c}\vec{v}^{\,q} \times\vec{B} \right)
\label{lorentzforcelaw}
\ .
\end{align}
The first four lines are Maxwell's equations and the last is the Lorentz force law.  Using these laws (and integration by parts, dropping boundary terms), we can derive an expression for local conservation of energy,
\begin{equation}
\vec{f}\cdot\vec{v}^{\,q}+\frac{\partial}{\partial t}\left(\frac{E^2}{8 \pi}+\frac{B^2}{8 \pi}\right)+\vec{\nabla}\cdot \left(\frac{c}{4 \pi} \vec{E} \times \vec{B}\right)=0
\ .
\label{energyconsEM}
\end{equation}
This is known as Poynting's theorem.\footnote{See \cite[ch.\ 27]{feynman2}; \cite[sec.\ 31]{landaulifshitzfields}; \cite[ch.\ 5]{hunt1991}; \cite[sec.\ 6.7]{jackson}; \cite[ch. 5]{lange}; \cite[ch.\ 3]{frisch2005}; \cite[sec.\ 15.4]{zangwill2012}; \cite[sec.\ 8.1.2]{griffiths}.}  Integrating \eqref{energyconsEM} over all of space, the third term drops out and we have an expression for global conservation of energy,
\begin{equation}
\int \vec{f} \cdot \vec{v}^{\,q}\ d^3 x\ +\ \frac{d}{d t} \int \left(\frac{E^2}{8 \pi}+\frac{B^2}{8 \pi}\right) d^3 x=0
\ ,
\end{equation}
that resembles \eqref{econservation1}, with an energy density for the electromagnetic field of
\begin{equation}
\rho^{\mathcal{E}}=\frac{E^2}{8 \pi}+\frac{B^2}{8 \pi}
\ .
\label{emfieldenergydensity}
\end{equation}
Integrating \eqref{energyconsEM} over a finite volume, the third term can be written as a surface integral (using the divergence theorem) and $\frac{c}{4 \pi} \vec{E} \times \vec{B}$ (the Poynting vector) can be interpreted as an energy flux density describing the flow of electromagnetic field energy across the surface of the volume.  There is much that could be said about the flow of energy, but it will not be our focus here.  We are concerned primarily with energy density, not energy flux density.

In electromagnetism, the electric field cannot generally be written as the gradient of a scalar potential, as in \eqref{epotential}, because it may have a non-zero curl \eqref{faraday}.  Using both a scalar potential $\phi$ and a vector potential $\vec{A}$, the electric and magnetic fields can be written as
\begin{align}
\vec{E}&=-\vec{\nabla}\phi - \frac{1}{c} \frac{\partial \vec{A}}{\partial t}
\label{potentials1}
\\
\vec{B}&=\vec{\nabla} \times \vec{A}
\ .
\label{potentials2}
\end{align}
In electrostatics, the move from a field energy density  \eqref{econservation1} to a potential energy density (or interaction energy density) \eqref{econservation2} is achieved by writing the electric field in terms of the scalar potential \eqref{epotential}, performing integration by parts, and then using \eqref{egauss} and \eqref{epotential} to rewrite the Laplacian of the potential in terms of the charge density.  This kind of maneuver cannot be pulled off in full electromagnetism, somehow using \eqref{potentials1} and \eqref{potentials2} to move from the field energy density in \eqref{emfieldenergydensity} to a potential energy density.

A brief aside:  If we restrict our attention to situations where the electric field is constant, then the magnetic contribution to the total field energy can be rewritten using \eqref{ampere} and \eqref{potentials2} as
\begin{align}
\int \frac{B^2}{8 \pi} d^3 x &= \int \left[\frac{1}{8 \pi} \vec{A}\cdot \left(\vec{\nabla} \times \left( \vec{\nabla} \times \vec{A}\right)\right)-\vec{\nabla}\cdot\left(\left(\vec{\nabla} \times \vec{A}\right)\times \vec{A}\right)\right] d^3 x
\nonumber
\\
&=\int \frac{1}{2c}\vec{J}\cdot\vec{A}\ d^3 x
\ ,
\label{magnetostaticenergy}
\end{align}
where the boundary term has been dropped in the move from the first line to the second.  The gauge-dependent expression $\frac{1}{2c}\vec{J}\cdot\vec{A}$ for the magnetic energy density can be found in textbook discussions of magnetostatics.  This form makes the magnetic energy look like a potential energy (dependent on the states of both matter and field), but it is only available in the rare circumstances where the electric field is constant.  We will encounter a similar expression in the next section.

Having seen a way of deriving the electromagnetic field's energy density from Maxwell's equations and the Lorentz force law in Poynting's theorem \eqref{energyconsEM}, let us now see (in sketch) how the same expression can be derived from the Lagrangian density.\footnote{My notation in this explanation of Lagrangian densities and energy-momentum tensors matches \cite[ch.\ 12]{jackson}, including the use of a $(+,-,-,-)$ signature for the metric.}  The standard Lagrangian density\footnote{This Lagrangian density is discussed in \cite[pg.\ 102]{barut1964}; \cite[ch.\ 8]{soper1976}; \cite[pg.\ 349]{weinbergQFT}; \cite[sec.\ 12.7]{jackson}; \cite[sec.\ 4.9]{rohrlich2007}; \cite[ch.\ 24]{zangwill2012}.} for the electromagnetic field interacting with some as-yet-unspecified matter is
\begin{equation}
\mathscr{L}=\mathscr{L}_m+\mathscr{L}_f-\frac{1}{c}J_{\mu}A^{\mu}
\ ,
\label{totallagrangian}
\end{equation}
where the final term describes the interaction of field and matter, the first term is the Lagrangian density for matter alone, and the second term is the Lagrangian density for the free electromagnetic field,
\begin{equation}
\mathscr{L}_f = -\frac{1}{16 \pi}F_{\mu\nu}F^{\mu\nu}=\frac{E^2}{8 \pi}-\frac{B^2}{8 \pi}
\label{fieldlagrangian}
\ .
\end{equation}
In \eqref{fieldlagrangian}, $F^{\mu\nu}$ is the Faraday tensor,
\begin{equation}
F^{\mu\nu}=\partial^\mu A^\nu - \partial^\nu A^\mu
\ ,
\end{equation}
where $A^\mu$ is the four-potential $(\phi,\vec{A})$---a combination of the scalar and vector potentials in \eqref{potentials1} and \eqref{potentials2}.  In \eqref{totallagrangian}, $J^{\mu}$ is the four-current $(c \rho^q,\vec{J}\,)$.

From the Lagrangian density in \eqref{fieldlagrangian}, one can derive an energy-momentum tensor for the electromagnetic field that gives energy and momentum densities for the electromagnetic field, as well as flux densities describing the flow of energy and momentum.  We will look at two candidates for such a tensor: the problematic but simply defined ``canonical'' energy-momentum tensor and the improved ``symmetric'' or ``symmetrized'' energy momentum tensor (with energy and energy flux densities matching those in Poynting's theorem).

The canonical energy-momentum tensor for the electromagnetic field can be calculated from the field Lagrangian density \eqref{fieldlagrangian} via
\begin{equation}
T_C^{\alpha \beta} = \frac{\partial \mathscr{L}_f}{\partial (\partial_\alpha A^\lambda)}\partial^\beta A^\lambda - g^{\alpha \beta}\mathscr{L}_f
\ ,
\label{canonicaldef}
\end{equation}
where here $g^{\alpha \beta}$ is the Minkowksi metric and
\begin{equation}
\frac{\partial \mathscr{L}_f}{\partial (\partial_\alpha A^\lambda)} = \frac{-1}{4 \pi}g^{\alpha \mu}F_{\mu \lambda}
\ .
\end{equation}
The $00$ component of the canonical energy-momentum tensor gives a candidate energy density for the electromagnetic field,
\begin{equation}
T_C^{00} = \frac{E^2}{8 \pi} + \frac{B^2}{8 \pi} + \frac{1}{4 \pi}\vec{E}\cdot\del \phi
\ .
\label{canonicalenergydensity}
\end{equation}
In the temporal gauge ($\phi=0$) this reduces to the previously derived field energy density in \eqref{emfieldenergydensity},\footnote{See \cite{creutz1979}.} but in general it will be different.  In the free case (without charged matter), the difference between \eqref{emfieldenergydensity} and \eqref{canonicalenergydensity} is a pure divergence that does not contribute to the total electromagnetic energy,
\begin{equation}
T_C^{00}=\frac{E^2}{8 \pi}+\frac{B^2}{8 \pi}+\frac{1}{4 \pi} \del \cdot (\phi \vec{E})
\end{equation}
by Gauss's law \eqref{gauss} with $\rho^q=0$.  When matter is present, the difference is important.  We have already seen in \eqref{energyconsEM} that the total energy of matter and field is conserved if we attribute the energy density in \eqref{emfieldenergydensity} to the electromagnetic field.  One might worry that the additional term in \eqref{canonicalenergydensity}, $\frac{1}{4 \pi}\vec{E}\cdot\del \phi$, would spoil conservation of energy.  In fact, energy will still be conserved provided that we also include an electromagnetic potential energy density of matter (or interaction energy density of matter and field) of $\rho^q \phi$.  Once we fill in a more detailed model of matter, this additional energy density should be derivable from \eqref{totallagrangian} as part of the complete canonical energy-momentum tensor for matter and field (as Soper \cite[sec.\ 8.3]{soper1976} shows for a charged fluid model of matter and as we will discuss for a Dirac field model of matter \eqref{totalcanonicalenergydensity}\footnote{The canonical energy-momentum tensor for the Dirac and electromagnetic fields will contain this $\rho^q \phi$ potential energy density, though that will not be the full potential energy density of matter.  An explanation as to why the other term does not lead to a violation of conservation of energy will be given near the end of section \ref{MDsection}.}).  With the $\rho^q \phi$ term included, the total electromagnetic energy is
\begin{equation}
\int \left( \frac{E^2}{8 \pi} + \frac{B^2}{8 \pi} + \frac{1}{4 \pi}\vec{E}\cdot\del \phi + \rho^q \phi\right)d^3 x
\label{correctedEMenergy}
\ .
\end{equation}
Using \eqref{gauss} to rewrite the charge density in terms of the electric field and integration by parts to move the spatial derivatives from $\vec{E}$ to $\phi$, the fourth term cancels the third and the total electromagnetic energy is the same as would be calculated from \eqref{emfieldenergydensity}.  Thus, as in Poynting's theorem, we would have conservation of total energy.

The canonical energy-momentum tensor is not regarded as an appealing energy-momentum tensor for the electromagnetic field and some authors skip over it entirely, moving directly to the symmetric energy-momentum tensor below.  When the canonical energy-momentum tensor is discussed, a number of shortcomings are listed to explain why it should not be taken to accurately describe the energy and momentum the electromagnetic field:  The components of the canonical energy-momentum tensor depend on the choice of gauge \cite[pg.\ 118]{barut1964}; \cite[pg.\ 117]{soper1976}; \cite[pg.\ 341]{doughty1990}; \cite[pg.\ 608]{jackson}.  The canonical energy-momentum tensor does not accurately capture the electromagnetic field's role as a source for gravitation in general relativity \cite[pg.\ 117]{soper1976}.  The canonical energy-momentum tensor is not symmetric \cite[pg.\ 118]{barut1964}; \cite[pg.\ 81]{landaulifshitzfields}; \cite[pg.\ 341]{doughty1990}.  Jackson \cite[pg.\ 608]{jackson} identifies this asymmetry as a ``drawback'' and writes that ``conservation of angular momentum requires that $T^{\alpha \beta}$ be symmetric,'' whereas Soper \cite[pg.\ 112--120]{soper1976} adds a spin angular momentum density that allows a form of angular momentum conservation for the canonical energy-momentum tensor (see also \cite[ch.\ 13]{konopinski1981}).  Though not directly discussing the canonical energy-momentum tensor for the electromagnetic field, Misner, Thorne, and Wheeler \cite[pg.\ 141--142]{mtw1973} explain that mass-energy equivalence requires energy flux density to be equal to momentum density (times $c^2$),\footnote{See also \cite[box 8.3]{lange}.} and thus that for any energy-momentum tensor $T^{0i}$ must equal $T^{i0}$.  The canonical energy-momentum tensor violates this condition.

Confining our attention to electrodynamics, there is a strong case that can be made against the canonical energy-momentum tensor, though one could argue that the case is not decisive.  In general relativity, the canonical energy-momentum tensor is not viable.  You need a symmetric energy-momentum tensor to serve as a source for gravity and this condition rules out the asymmetric canonical energy-momentum tensor.  Appealing to general relativity, a different energy-momentum tensor---the symmetric energy momentum tensor---can be derived from the Lagrangian density of the electromagnetic field on curved spacetime by taking a variational derivative of the action $\int \mathscr{L}_f \, d^4 x$ with respect to the metric and then considering the special case of flat spacetime (as described in \cite[pg.\ 195--196]{soper1976}; \cite[sec.\ 1.2 and 9.1]{wald2022}).  Combining the considerations that come from within electrodynamics and from an eye peeking ahead towards general relativity, let us set aside the canonical energy-momentum tensor.

We can move from the problematic canonical energy-momentum tensor to the symmetric (or ``symmetrized'') energy-momentum tensor for the electromagnetic field by adding an additional term,
\begin{align}
T^{\alpha \beta} &= T_c^{\alpha \beta}- \frac{1}{4 \pi}\partial_\lambda (F^{\lambda \alpha} A^\beta)
\nonumber
\\
&= \frac{1}{4 \pi}g^{\alpha \mu}F_{\mu \lambda}F^{\lambda \beta}  + \frac{1}{16 \pi}g^{\alpha\beta}F_{\mu \lambda}F^{\mu \lambda}
\ ,
\label{sym}
\end{align}
from which it follows that $T^{\alpha \beta} = T^{\beta \alpha}$.  It is this tensor that is generally taken to accurately describe the densities and flows of energy and momentum in the electromagnetic field.  Although we will not go through the details here, the additional term that was added without explanation in \eqref{sym} can be derived (either from general relativity, as was mentioned in the previous paragraph, or without appealing to general relativity\footnote{See \cite{belinfante1939}; \cite{rosenfeld1940}; \cite[appendix 1]{wentzel1949}; \cite[ch.\ 3]{barut1964}; \cite{goedecke1973}; \cite[sec.\ 9.5]{soper1976}; \cite[sec.\ 19.5]{doughty1990}; \cite[sec.\ 7.4]{weinbergQFT}; \cite[pg.\ 47--49]{greiner1996}; \cite[ch.\ 7]{low1997}; \cite{baker2022}.}).

The field energy density and energy flux density in the symmetric energy-momentum tensor are exactly the same as the ones in Poynting's theorem \eqref{energyconsEM}:
\begin{align}
T^{00} &= \frac{E^2}{8 \pi} + \frac{B^2}{8 \pi}
\\
c T^{0i} &= \frac{c}{4 \pi} (\vec{E} \times \vec{B})_i
\ .
\end{align}
At this point, potential energy seems to have disappeared.  The above electromagnetic energy and energy flux densities are sufficient to ensure conservation of energy in interactions between matter and field, balancing the $\vec{f} \cdot \vec{v}^{\,q}$ rate at which electromagnetic forces do work on matter per unit volume.  There is no apparent need for potential energy or interaction energy in classical electromagnetism.\footnote{Soper \cite[pg.\ 121]{soper1976} remarks on a few ``nice things'' that happen when one moves from the canonical to the symmetric electromagnetic energy-momentum tensor, including the fact that (for a particular charged fluid model of matter) the total energy-momentum tensor for matter and field can be written as the sum of a separate matter energy-momentum tensor and field energy-momentum tensor such that ``one may speak of the energy of matter ... and the energy of the electromagnetic field ... without needing an interaction energy like [$\rho^q \phi$].''}

Before proceeding to the reappearance of potential energy in the next section, let me mention an interesting way that the energy of the electromagnetic field can be rewritten in the Coulomb gauge.\footnote{Here I follow moves made in \cite[sec.\ 15.2]{bjorkendrellfields}; \cite[sec.\ 8.1]{hatfield}.}  From \eqref{emfieldenergydensity} and \eqref{potentials2}, the energy of the electromagnetic field can be written in terms of the vector and scalar potentials as
\begin{equation}
\int \left( \frac{E^2}{8 \pi} + \frac{B^2}{8 \pi}\right)d^3 x=\int \Bigg( \frac{|\vec{\nabla}\phi|^2+\frac{2}{c}\vec{\nabla}\phi\cdot\frac{\partial \vec{A}}{\partial t}+\frac{1}{c^2} \left|\frac{\partial \vec{A}}{\partial t}\right|^2}{8 \pi} + \frac{(\vec{\nabla}\times\vec{A})^2}{8 \pi}\Bigg)d^3 x
\ .
\end{equation}
The middle term in the expansion of $\frac{E^2}{8 \pi}$ can be eliminated by using integration by parts to move the spatial derivatives to the vector potential, where the Coulomb gauge condition ($\vec{\nabla}\cdot\vec{A}=0$) allows us to drop the term.  Integration by parts and Gauss's law \eqref{gauss} can then be used to rewrite the first term in the expansion of $\frac{E^2}{8 \pi}$,
\begin{align}
\int \left( \frac{E^2}{8 \pi} + \frac{B^2}{8 \pi}\right)d^3 x&=\int \Bigg( \frac{-\phi \nabla^2 \phi+\frac{1}{c^2} \left|\frac{\partial \vec{A}}{\partial t}\right|^2}{8 \pi} + \frac{(\vec{\nabla}\times\vec{A})^2}{8 \pi}\Bigg)d^3 x
\nonumber
\\
&=\int \Bigg( \frac{1}{2} \rho^q \phi+\frac{\left|\frac{\partial \vec{A}}{\partial t}\right|^2}{8 \pi c^2}  + \frac{(\vec{\nabla}\times\vec{A})^2}{8 \pi}\Bigg)d^3 x
\ .
\label{returntooldPE}
\end{align}
The first term in this expression is precisely the potential energy density from electrostatics \eqref{epotentialenergydensity} (illustrated for a sphere of uniform charge on the right side of figure \ref{energyfigure}).  As in \eqref{epotentialenergydensity2}, \eqref{returntooldPE} can be written as
\begin{equation}
\int \left( \frac{E^2}{8 \pi} + \frac{B^2}{8 \pi}\right)d^3 x=\int \Bigg( \frac{1}{2}\int{\frac{\rho^q(\vec{x}) \rho^q(\vec{x}')}{|\vec{x}-\vec{x}'|}\, d^3 x'}+\frac{\left|\frac{\partial \vec{A}}{\partial t}\right|^2}{8 \pi c^2} + \frac{(\vec{\nabla}\times\vec{A})^2}{8 \pi}\Bigg)d^3 x
\ .
\label{returntooldPE2}
\end{equation}
Appealing to the above expressions, one might argue that the potential energy density $\frac{1}{2} \rho^q \phi$ from electrostatics is viable in electromagnetism provided one endows the electromagnetic field with a field energy density of $\frac{\left|\frac{\partial \vec{A}}{\partial t}\right|^2}{8 \pi c^2}  + \frac{(\vec{\nabla}\times\vec{A})^2}{8 \pi}$.  Although the above expressions may be useful for certain purposes, there are reasons to resist treating $\frac{1}{2} \rho^q \phi$ as a density of potential energy in classical electromagnetism:  First, the above maneuvers required adopting a particular gauge (whereas \eqref{emfieldenergydensity} gave a gauge-invariant energy density for the electromagnetic field).  Second, it is unclear how one would incorporate the new potential energy and electromagnetic field energy densities into energy-momentum tensors for matter and the electromagnetic field.  After constructing such energy-momentum tensors, one could compare their virtues to the canonical and symmetric energy-momentum tensors.

\section{Classical Maxwell-Dirac Field Theory}\label{MDsection}

In the previous section, we left the Lagrangian density of matter unspecified.  In preparation for our discussion of quantum electrodynamics in the next section, we can take this ``matter'' to be another field: the four-component complex-valued\footnote{Because the Dirac field operators anticommute in quantum field theory, the classical Dirac field is sometimes treated as Grassmann-valued instead of complex-valued. For more on this issue, see \cite[appendix A]{positrons}; \cite[sec.\ 5.1]{fundamentalityoffields} and references therein.} Dirac field $\psi$.  Let us thus consider a classical theory of interacting electromagnetic and Dirac fields, sometimes called ``Maxwell-Dirac field theory.''  This classical field theory does not arise as a classical limit to quantum field theory,\footnote{See \cite[pg.\ 444, 453]{doughty1990}; \cite[pg.\ 221]{duncan}.} but it is still worthy of study because, upon field quantization, it yields quantum electrodynamics (the quantum field theory that describes interactions between photons, electrons, and positrons).

The Lagrangian density describing classical interactions between the Dirac and electromagnetic fields is the total Lagrangian density from the previous section\footnote{Note that for the theory of interacting scalar and electromagnetic fields that is the classical precursor to scalar quantum electrodynamics (as contrasted with spinor quantum electrodynamics, which will be the quantum successor of the classical field theory examined here), the Lagrangian density does not fit the general form in \eqref{totallagrangian} as the interaction is different.  (See \cite[eq.\ 9.11]{schwartz} for the Lagrangian density of scalar quantum electrodynamics.)} \eqref{totallagrangian} with $\mathscr{L}_f$ given in \eqref{fieldlagrangian}, $J^{\mu}$ given by the four-current density of the Dirac field,
\begin{equation}
J^{\mu}=-e c \psi^\dagger \gamma^0 \gamma^\mu \psi
\label{diraccurrent}
\end{equation}
(where $-e$ is the charge of the electron), and $\mathscr{L}_m$ given by either
\begin{equation}
\mathscr{L}_m=i \hbar c \, \psi^\dagger \gamma^0\gamma^\mu\partial_\mu \psi - m c^2 \psi^\dagger \gamma^0 \psi
\label{lagrangian1}
\end{equation}
or\footnote{These two Lagrangian densities differ by a four-divergence and yield the same field equations \cite[pg.\ 102--103]{barut1964}; \cite[pg.\ 221]{grandy1991}; \cite[pg.\ 117--122]{greiner1996}.  The first is used in \cite{wentzel1949, heitler1954, bjorkendrellfields, hatfield} and the second is used in \cite{schwinger1948, schweber1961, doughty1990, ryder1996}.}
\begin{equation}
\mathscr{L}_m=\frac{1}{2}\psi^\dagger\left(i \hbar c \, \gamma^0\gamma^\mu\partial_\mu -m c^2 \right)\psi+\frac{1}{2}\left(i \hbar c \, (\partial_\mu \psi^\dagger) \gamma^0\gamma^\mu- m c^2\right)\psi
\ ,
\label{lagrangian2}
\end{equation}
where the $\partial_\mu$ acting on $\psi^\dagger$ acts only on $\psi^\dagger$.  From the total Lagrangian density \eqref{totallagrangian}, one can derive\footnote{See \cite[pg.\ 132--134]{heitler1954}; \cite[pg.\ 144]{barut1964}; \cite[pg.\ 275--276]{schweber1961}; \cite[pg.\ 451]{doughty1990}.} the Maxwell-Dirac equations: Maxwell's equations\footnote{To be precise, two of Maxwell's equations are derived from the Lagrangian density and two are automatically satisfied by the use of scalar and vector potentials satisfying \eqref{potentials1} and \eqref{potentials2}.} \eqref{gauss}--\eqref{ampere} and the Dirac equation,
\begin{equation}
i \hbar \frac{\partial \psi}{\partial t}=\big(-i \hbar c\, \gamma^0\vec{\gamma}\cdot\vec{\nabla} + \gamma^0 m c^2-e \gamma^0 \gamma_\mu A^\mu \big)\psi
\ .
\label{dirac}
\end{equation}
Sometimes $\gamma^0\vec{\gamma}$ is written as $\vec{\alpha}$ and $\gamma^0$ as $\beta$, but we will not use that notation here.

Within this classical theory of interacting Dirac and electromagnetic fields, gauge transformations take the form
\begin{align}
A^\mu &\rightarrow A^\mu + \partial^\mu \alpha
\nonumber
\\
\psi &\rightarrow e^{- i \frac{e}{\hbar c} \alpha} \psi
\ ,
\label{gaugetransformations}
\end{align}
where $\alpha$ can be any scalar function of space and time.\footnote{See \cite[pg.\ 1442]{schwinger1948}; \cite[pg.\ 477--478]{creutz1979}; \cite[sec.\ 20.9]{doughty1990}; \cite[pg.\ 151]{greiner1996}; \cite[pg.\ 57]{nakahara2003}.}  Note that these transformations involve $\psi$ as well as $\phi$ and $\vec{A}$, whereas in textbook treatments of classical electromagnetism we generally assume that gauge transformations act only on the scalar and vector potentials (not on the matter interacting with these potentials).

If we apply the same recipe that was used to generate the canonical energy-momentum tensor of the electromagnetic field to the full Lagrangian density for interacting Dirac and electromagnetic fields \eqref{totallagrangian}, including $\mathscr{L}_m$ from \eqref{lagrangian1},\footnote{For discussion of the canonical energy-momentum tensor generated from $\mathscr{L}_m$ in \eqref{lagrangian2}, see \cite[pg.\ 1444]{schwinger1948}; \cite[pg.\ 170, footnote 1]{wentzel1949}; \cite[pg.\ 440--441]{doughty1990}; \cite[pg.\ 121--122]{greiner1996}.} we get a total canonical energy-momentum tensor of
\begin{equation}
T_C^{\alpha \beta} = \frac{\partial \mathscr{L}}{\partial (\partial_\alpha A^\lambda)}\partial^\beta A^\lambda+\frac{\partial \mathscr{L}}{\partial (\partial_\alpha \psi^\lambda)}\partial^\beta \psi^\lambda - g^{\alpha \beta}\mathscr{L}
\ .
\label{totalcanonicaldef}
\end{equation}
The $00$ component is a candidate energy density for the two interacting fields,\footnote{This energy density is discussed in \cite[pg.\ 171]{wentzel1949}; \cite[pg.\ 132, 419]{heitler1954}; \cite[eq.\ 15.13]{bjorkendrellfields}; \cite[eq.\ 8.7]{hatfield}; \cite{leclerc2006}.}
\begin{align}
T_C^{00} &=\frac{E^2}{8 \pi} + \frac{B^2}{8 \pi} + \frac{1}{4 \pi}\vec{E}\cdot\del \phi+i\hbar \psi^\dagger \frac{\partial \psi}{\partial t}
\nonumber
\\
&=\frac{E^2}{8 \pi} + \frac{B^2}{8 \pi} + \frac{1}{4 \pi}\vec{E}\cdot\del \phi
+\psi^\dagger\big(-i \hbar c\, \gamma^0\vec{\gamma}\cdot\vec{\nabla} + \gamma^0 m c^2\big)\psi-e\psi^\dagger\gamma^0 \gamma_\mu \psi A^\mu
\ .
\label{totalcanonicalenergydensity}
\end{align}
If this were the correct energy density for the two fields, the last term in the second line might be interpreted as a potential energy density for the Dirac field or an interaction energy density\footnote{Discussing interactions between the electromagnetic field and matter in general, Konopinski \cite[pg.\ 419, 425]{konopinski1981} identifies this kind of term as an interaction energy density.} for the two fields together.  However, given the problems for the canonical energy-momentum tensor that were recounted in the previous section, we should be skeptical as to whether \eqref{totalcanonicalenergydensity} gives the correct energy density for the two fields.  Continuing with this problematic energy density for a moment, we can calculate the total energy by integrating over all of space.  When we do so, the contributions from the $\frac{1}{4 \pi}\vec{E}\cdot\del \phi$ and $-e\psi^\dagger\gamma^0 \gamma_0 \psi A^0=\rho^q \phi$ terms cancel (as can be seen by using integration by parts to move the spatial derivatives from $\phi$ to $\vec{E}$ and Gauss's law \eqref{gauss} to rewrite $\vec{\nabla}\cdot \vec{E}$ in terms of $\rho^q=-e \psi^\dagger \psi$):\footnote{See \cite[eq.\ 15.14]{bjorkendrellfields}; \cite[eq.\ 3.6]{creutz1979}; \cite[eq.\ 8.9]{hatfield}.}
\begin{equation}
\mathcal{E}=\int \left( \frac{E^2}{8 \pi} + \frac{B^2}{8 \pi} +\psi^\dagger\big(-i \hbar c\, \gamma^0\vec{\gamma}\cdot\vec{\nabla} + \gamma^0 m c^2\big)\psi+ e\, \psi^\dagger\gamma^0\vec{\gamma}\psi \cdot \vec{A}\right)d^3 x
\ .
\label{canonicaltotalenergy}
\end{equation}
Although the canonical energy-momentum tensor has unappealing features, we will see shortly that the above total energy agrees with the total energy calculated from the corrected symmetric energy-momentum tensor (so we can take \eqref{canonicaltotalenergy} to be the Hamiltonian for this classical field theory).

Before discussing the symmetric energy-momentum tensor, let me mention that in the Coulomb gauge ($\vec{\nabla}\cdot \vec{A}=0$) one can write the total energy \eqref{canonicaltotalenergy} as in \eqref{returntooldPE2},\footnote{This expression \eqref{canonicaltotalenergycoulomb} appears in \cite[sec.\ 15.2]{bjorkendrellfields}; \cite[sec.\ 8.3]{weinbergQFT}; \cite[pg.\ 983]{leclerc2006}; \cite[sec.\ 6.4]{tong} and a similar expression appears in \cite[sec.\ 8.1]{hatfield}.}
\begin{align}
\mathcal{E}&=\int \left( \frac{\big|\frac{\partial\vec{A}}{\partial t}\big|^2}{8 \pi c^2}
 + \frac{(\vec{\nabla}\times \vec{A})^2}{8 \pi} +\psi^\dagger\big(-i \hbar c\, \gamma^0\vec{\gamma}\cdot\vec{\nabla} + \gamma^0 m c^2\big)\psi + e\, \psi^\dagger\gamma^0\vec{\gamma}\psi \cdot \vec{A}+ \frac{1}{2} \int \frac{\rho^q(\vec{x})\rho^q(\vec{x}')}{|\vec{x}-\vec{x}'\,|} d^3 x'\right)d^3 x
\ .
\label{canonicaltotalenergycoulomb}
\end{align}
As in our discussion of \eqref{returntooldPE2}, one might consider interpreting the second-to-last term in \eqref{canonicaltotalenergycoulomb} as a potential energy of the Dirac field and the last term as a distinct potential energy of the Dirac field (leaving the first two terms in \eqref{canonicaltotalenergycoulomb} to be interpreted as giving a new energy density for the electromagnetic field).  This alternative analysis in the Coulomb gauge is interesting, but would not yield gauge-invariant expressions for the energy densities of the electromagnetic and Dirac fields.  Also, it is unclear how the proposed energy densities would be incorporated into energy-momentum tensors for the two fields.  Still, \eqref{canonicaltotalenergycoulomb} is a useful way of rewriting the total energy \eqref{canonicaltotalenergy} that sometimes appears in quantum field theory textbooks.

The symmetric energy-momentum tensor for the interacting electromagnetic and Dirac fields is the sum of the symmetric energy-momentum tensor for the electromagnetic field in \eqref{sym}, labeled $T_f^{\alpha \beta}$, and an energy-momentum tensor for the Dirac field, labeled $T_m^{\alpha \beta}$,
\begin{align}
&T^{\alpha \beta} = \overbrace{\frac{1}{4 \pi}g^{\alpha \mu}F_{\mu \lambda}F^{\lambda \beta} + \frac{1}{16 \pi}g^{\alpha\beta}F_{\mu \lambda}F^{\mu \lambda}}^{\mbox{$T_f^{\alpha \beta}$}}
\nonumber
\\
&+ \underbrace{\frac{i\hbar c}{4}\Big(\psi^\dagger\gamma^0 \gamma^\alpha \partial^\beta  \psi + \psi^\dagger\gamma^0 \gamma^\beta \partial^\alpha  \psi - (\partial^\beta \psi^\dagger)\gamma^0 \gamma^\alpha \psi - (\partial^\alpha \psi^\dagger)\gamma^0 \gamma^\beta \psi\Big)
+\frac{1}{2}\psi^\dagger\left(e \gamma^0 \gamma^\alpha A^\beta+e \gamma^0 \gamma^\beta A^\alpha\right)\psi}_{\mbox{$T_m^{\alpha \beta}$}}
\ ,
\label{fullsymtensor}
\end{align}
where the derivatives acting on $\psi^\dagger$ are understood to act only on $\psi^\dagger$, as in \eqref{lagrangian2}.\footnote{The symmetric energy-momentum tensor for interacting Dirac and electromagnetic fields is given in \cite[pg.\ 1444]{schwinger1948}; \cite[sec.\ 21]{wentzel1949}; \cite[pg.\ 419]{heitler1954}; \cite{goedecke1973, inglis2016}.}  In terms of the gauge covariant derivative $D^\mu = \partial^\mu + i \frac{e}{\hbar c} A^\mu$, the energy-momentum tensor for the Dirac field can be written as
\begin{equation}
T_m^{\alpha \beta}=\frac{i\hbar c}{4}\Big(\psi^\dagger\gamma^0 \gamma^\alpha D^\beta  \psi + \psi^\dagger\gamma^0 \gamma^\beta D^\alpha  \psi - (D^{*\beta} \psi^\dagger)\gamma^0 \gamma^\alpha \psi - (D^{*\alpha} \psi^\dagger)\gamma^0 \gamma^\beta \psi\Big)
\ ,
\label{diracenergymomentum}
\end{equation}
where $D^{*\mu} = \partial^\mu - i \frac{e}{\hbar c} A^\mu$.  Because $D^\mu \psi \rightarrow e^{-i \frac{e}{\hbar c} \alpha}D^\mu \psi$ under the gauge transformation in \eqref{gaugetransformations}, the above form of $T_m^{\alpha \beta}$ makes it clear that the components of the tensor are invariant under gauge transformations.  Setting aside interactions with the electromagnetic field, the gauge covariant derivatives in \eqref{diracenergymomentum} can be replaced by ordinary derivatives to arrive at the symmetric energy-momentum tensor for the free Dirac field.  Going the other direction, the energy-momentum tensor in \eqref{diracenergymomentum} can be found from the symmetric energy-momentum tensor for the free Dirac field by replacing the ordinary derivatives with gauge covariant derivatives.

The $00$ component of $T_m^{\alpha \beta}$ gives a gauge-invariant energy density for the Dirac field,
\begin{align}
T_m^{00}&= \frac{i\hbar}{2} \psi^{\dagger}\frac{\partial \psi}{\partial t} -\frac{i\hbar}{2} \frac{\partial \psi^{\dagger}}{\partial t}\psi + e \phi \psi^\dagger \psi
\nonumber
\\
&= \psi^{\dagger}\left(\gamma^0 m c^2\right)\psi \underbrace{- \frac{i \hbar c}{2}\left(\psi^{\dagger} \gamma^0\vec{\gamma}\cdot\del \psi
 - (\del \psi^{\dagger}) \cdot (\gamma^0\vec{\gamma} \psi )\right)}_{\mbox{(A)}}+ \underbrace{e\, \psi^\dagger\gamma^0\vec{\gamma}\psi \cdot \vec{A}}_{\mbox{(B)}}
 \ .
 \label{diracenergydensity}
\end{align}
Integrating $T_f^{00}$ and $T_m^{00}$ over all of space and using integration by parts to move the gradient from $\psi^{\dagger}$ to $\psi$, we see that the total energy is the same as for the canonical energy-momentum tensor \eqref{canonicaltotalenergy}.\footnote{This point has been noted for the free Dirac field in \cite[pg.\ 171]{wentzel1949}; \cite[pg.\ 419]{heitler1954}; \cite[pg.\ 219]{schweber1961}; \cite[footnote 8]{positrons}.}  The terms labeled (A) and (B) in \eqref{diracenergydensity} can be combined using the gauge covariant derivative to rewrite the energy density as
\begin{equation}
T_m^{00}= \psi^{\dagger}\left(\gamma^0 m c^2\right)\psi \underbrace{- \frac{i \hbar c}{2}\left(\psi^{\dagger} \gamma^0 \gamma^i D^i \psi
 - (D^{* i} \psi^{\dagger}) \gamma^0 \gamma^i \psi \right)}_{\mbox{(A)+(B)}}
 \ .
  \label{diracenergydensity2}
\end{equation}

The (B) term,
\begin{equation}
e\, \psi^\dagger\gamma^0\vec{\gamma}\psi \cdot \vec{A}=-\frac{1}{c}\vec{J}\cdot \vec{A}
\ ,
\label{magform}
\end{equation}
looks like it could be a potential energy for the Dirac field or an interaction energy between the Dirac and electromagnetic fields.\footnote{There is an alternative picture of potential energy available.  Focusing on the first line of \eqref{diracenergydensity}, we might interpret $e \phi \psi^\dagger \psi = - \rho^q \phi$ as the potential energy of the Dirac field (which disappears in the temporal gauge) and the rest of that line as the non-potential energy of the Dirac field.  On that interpretation, the Dirac field's non-potential energy density depends on the field's rate of change---unlike the energy density of the electromagnetic field \eqref{emfieldenergydensity}.}  This term has the same form as the magnetostatic energy density in \eqref{magnetostaticenergy}, though the sign is flipped and the magnitude doubled.  Like that energy density, (B) is gauge-dependent.  This defect seemed unavoidable for the electric and magnetic potential energy densities in \eqref{epotentialenergydensity} and \eqref{magnetostaticenergy}, but here the gauge transformation properties of the Dirac field \eqref{gaugetransformations} provide a remedy.  Although (B) is gauge-dependent, the combination (A)+(B) is gauge-invariant---as is clear from \eqref{diracenergydensity2}.  Because the remainder of the field's energy density, $\psi^{\dagger}\left(\gamma^0 m c^2\right)\psi$, is gauge-invariant, the total energy density of the Dirac field \eqref{diracenergydensity2} is gauge-invariant.  This helps us to settle a type of question that we had often left unanswered in the previous sections: whether a certain quantity should be interpreted as a potential energy density or an interaction energy density.  Classifying (B) as potential energy density places it together with the rest of the energy of matter, yielding a total energy density for the Dirac field that is gauge-invariant.  If we instead classify that term as an interaction energy density possessed by matter and field together, then we would have a gauge-dependent matter energy density and a separate gauge-dependent interaction energy density.  Thus, it is better to view (B) as a potential energy density of the Dirac field than as an interaction energy density of the two fields.  Another alternative would be to interpret (B) as part of the electromagnetic field's energy density, using one of Maxwell's equations \eqref{ampere} to replace the current density in \eqref{magform} and express (B) purely in terms of electromagnetic field variables.  This option is unattractive as it would make both the total energy density of the electromagnetic field and the total energy density of the Dirac field gauge-dependent.

Because (B) is gauge-dependent but (A)+(B) is not, one might be tempted to regard (A)+(B) as the potential energy density instead of (B) alone.  This alternative candidate potential energy density is discussed further in appendix \ref{KPED} (where the Aharonov-Bohm effect is considered).  Although there may be some lingering uncertainty as to whether (B) or (A)+(B) is the correct potential energy density, note that there is no question here as to whether the Dirac field carries potential energy, no question as to what the Dirac field's total energy density is at any point in space, and no question as to how that energy density depends on the state of the electromagnetic field.  Moving forward, we will take (B) to be the Dirac field's potential energy density.

It is natural to wonder which terms in \eqref{diracenergydensity} give the Dirac field's kinetic energy density, but I will not attempt to settle that question here.  One could explore taking (A), (A)+(B), or the entirety of $T_m^{00}$ to be the field's kinetic energy density.\footnote{Feynman \emph{et al.{}} \cite[sec.\ 21.1--21.3]{feynman3} include a discussion of two different kinds of momentum for classical and quantum particles that may be helpful in assessing whether (A) or (A)+(B) should be regarded as the kinetic energy density of the Dirac field.  Leclerc \cite[pg.\ 983]{leclerc2006} takes the integral of $T_m^{00}$ to give the total kinetic energy of the Dirac field.}  On two of these options, the potential energy density (B) would be part of the kinetic energy density.  If you think that a strong case can be made for one of those two options, you might challenge the idea that (B) should be interpreted as a potential energy density on the grounds that kinetic and potential energies must be distinct.  Although it is an unfamiliar idea, I am open to the possibility that the Dirac field's potential energy density might be part of its kinetic energy density.\footnote{If we take one of the two options where the Dirac field's potential energy (B) is part of its kinetic energy density, then the kinetic energy of the Dirac field will not be an intrinsic property of the field (contra footnote \ref{KEfootnote}).}

To better understand the Dirac field's potential energy density (B), let us focus for a moment on situations where the electric field is constant.  In such situations, the total potential energy of the field can be written as
\begin{equation}
\int -\frac{1}{c}\vec{J}\cdot\vec{A}\ d^3 x=\int -\frac{B^2}{4 \pi} d^3 x
\ ,
\end{equation}
via \eqref{magnetostaticenergy} and \eqref{magform}.  Using this form, the total energy of the Dirac and electromagnetic fields in \eqref{canonicaltotalenergy} can be written (for a constant electric field) as
\begin{equation}
\mathcal{E}=\int \left( \frac{E^2}{8 \pi} - \frac{B^2}{8 \pi} +\psi^\dagger\big(-i \hbar c\, \gamma^0\vec{\gamma}\cdot\vec{\nabla} + \gamma^0 m c^2\big)\psi \right)d^3 x
\ .
\end{equation}
Combining the Dirac field's potential energy with the electromagnetic field energy flips the sign of the $B^2$ term.  This means that when charge flows in the Dirac field, the magnetic field that is created lowers the total energy of matter and field.  As an example, this would happen for states of the Dirac field representing a single spinning electron\footnote{Such states are discussed in \cite{ohanian1986, howelectronsspin, smallelectronstates, spinmeasurement, bb2021}.} (if we assume that the electric field is constant, as might be the case in a classical model of the hydrogen atom where the electron's charge distribution is held in place by the electric field of the proton at the center of the atom).

Our above observation that the Dirac field possesses potential energy raises an important question: \emph{How is it possible that the Dirac field has a potential energy density given the absence of potential energy in the previous section?}  In classical electrodynamics, potential energy alone cannot be used to achieve conservation of energy.  By contrast, electromagnetic field energy can be used to achieve conservation of energy without any apparent need for potential energy.  Being precise, our analysis of Poynting's theorem showed that if we take $\vec{f} \cdot \vec{v}^{\,q}=\vec{J}\cdot\vec{E}$ to be the rate at which energy is transferred to matter by electromagnetic forces per unit volume, that energy transfer can be balanced by changes in the energy of the electromagnetic field---provided we endow the field with its standard energy density of $\frac{E^2}{8 \pi} + \frac{B^2}{8 \pi}$.  In an ordinary explanation of Poynting's theorem, Jackson \cite[pg.\ 258]{jackson} writes that the integral of $\vec{J}\cdot\vec{E}$ over all of space ``represents a conversion of electromagnetic energy into mechanical or thermal energy.''  But, we need not assume that the energy of matter is entirely non-electromagnetic.  We can understand $\vec{f} \cdot {v}^{\,q}$ as giving the rate at which electromagnetic field energy is transferred to matter per unit volume, allowing that the energy of that matter may include electromagnetic potential energy, as in \eqref{diracenergydensity}.

This moves potential energy from one side of the ledger to the other.  In electrostatics, we saw that energy can be conserved by introducing an electric potential energy that balances the energy changes induced by electric forces, with this potential energy falling when electric forces do positive work on matter and rising when electric forces do negative work \eqref{econservation2}.  In classical Maxwell-Dirac field theory, energy can be conserved by using the standard electromagnetic field energy to balance the energy transferred to matter by electromagnetic forces, with this field energy falling when electromagnetic forces do positive work on matter and rising when electric forces do negative work.  The charged matter (the Dirac field) has electromagnetic potential energy and the rise and fall of this potential energy will ensure that the total energy of matter increases and decreases appropriately as electromagnetic forces do work on matter.  By reconsidering the role of potential energy in balancing transfers of energy, we have arrived at an explanation as to how potential energy was able to reappear in this section after it seemed to have disappeared in the last.

Two asides before moving on to the next section:  First, let us consider the connection between energy and mass.  One way of understanding mass-energy equivalence is as the claim that anything with relativistic mass $m_r$ has energy $m_r c^2$ and anything with energy $\mathcal{E}$ has relativistic mass $\mathcal{E}/c^2$ \cite[appendix A]{gravitationalfield}.  For example, a body with rest mass $m_0$ traveling at some velocity $v$ has relativistic mass $m_0 \gamma$ and energy $m_0\gamma c^2$, where $\gamma =\frac{1}{\sqrt{1-\frac{v^2}{c^2}}}$.  The electromagnetic field has an energy density and, applying the above version of mass energy equivalence, it also has a relativistic mass density equal to its energy density divided by $c^2$ \cite[box 8.3]{lange}; \cite{forcesonfields, gravitationalfield}.  Similarly, the Dirac field has a relativistic mass density equal to its energy density divided by $c^2$ \cite[pg.\ 238]{takabayasi1956}; \cite[pg.\ 35]{takabayasi1957}; \cite{howelectronsspin, smallelectronstates}.  We have seen here that this mass of the Dirac field is not an intrinsic property of the field.  It depends on the state of the electromagnetic field.

Second, note that the energy density for the classical Dirac field \eqref{diracenergydensity} that appears in the symmetric energy-momentum tensor can be either positive or negative, even when the scalar and vector potentials vanish.  In discussions of the Dirac equation, the problems involved with handling negative energies are infamous.  Elsewhere, I have argued that we ought to revise our classical theory of the free Dirac field so that both positive and negative frequency modes are associated with positive energies \cite{positrons}---though it is still possible for the energy density to be negative in certain regions \cite{bb2021}.  In this revised theory of the Dirac field, negative frequency modes are viewed as positive energy and positive charge positron modes.  The new energy density for the free Dirac field can be expressed in a form that resembles \eqref{diracenergydensity} \cite[footnote 19]{positrons}.  This revised classical field theory has some advantages: it immediately yields the correct quantum field theory for the free Dirac field upon quantization and avoids complications involving normal-ordering and dismissing infinite contributions to the energy and charge.  It would be valuable to explore modifying the classical theory of interacting Dirac and electromagnetic fields analyzed in this section along similar lines, including positive charge and revising the energy density to streamline the transition to quantum field theory.  However, it is not obvious how to best do so.  Thus, for our purposes here, we have stuck with the standard way of formulating a classical theory of interacting Dirac and electromagnetic fields where the Dirac field's charge is uniformly negative and negative frequency modes are associated with negative energy.

\section{Quantum Electrodynamics}\label{QFTsection}

The standard Hamiltonian for quantum electrodynamics\footnote{This Hamiltonian is given in \cite[eq.\ 15.14]{bjorkendrellfields}; \cite[eq.\ 3.6]{creutz1979}; \cite[eq.\ 8.9]{hatfield}; \cite[sec.\ 6.4]{tong}.  A variant is given in \cite[eq.\ 15]{birula1984}.} is
\begin{equation}
\widehat{H}=\int \left( \frac{\widehat{E}^2}{8 \pi} + \frac{\widehat{B}^2}{8 \pi} +\widehat{\psi}^\dagger\big(-i \hbar c\, \gamma^0\vec{\gamma}\cdot\vec{\nabla} + \gamma^0 m c^2\big)\widehat{\psi}+ e\, \widehat{\psi}^\dagger\gamma^0\vec{\gamma}\widehat{\psi} \cdot \hat{\vec{A}}\right)d^3 x
\ ,
\label{QEDhamiltonian}
\end{equation}
though this common way of writing it obscures some normal-ordering that will have to be done.  This Hamiltonian operator gives the evolution of the quantum state $| \Psi (t)\rangle$ via the Schr\"{o}dinger equation, in its general form
\begin{equation}
i \hbar \frac{d}{d t}| \Psi (t)\rangle=\widehat{H}| \Psi(t) \rangle
\ .
\label{SE}
\end{equation}
There is disagreement as to the nature of these quantum states, but here we will stay on common ground by focusing on operators.  The Hamiltonian \eqref{QEDhamiltonian} is the operator version of our classical total energy for the Dirac and electromagnetic fields in the previous section \eqref{canonicaltotalenergy}, where the potential energy appears at the end of the Hamiltonian as an interaction term.  The total energy operator \eqref{QEDhamiltonian} is the same whether one uses the canonical or the symmetric energy-momentum tensor.  Although the total energy operator is the same, the canonical and symmetric energy-momentum tensors yield different operators for energy density, momentum density, and momentum flux density.

\begin{figure}[htb]
\center{
\begin{fmffile}{vertex}
\begin{fmfgraph*}(125,100)
\fmfleft{a,c}
\fmfright{d}
\fmf{fermion}{a,b,c}
\fmf{photon}{b,d}
\end{fmfgraph*}
\end{fmffile}
}
\caption{The radiative interaction vertex that appears in Feynman diagrams for quantum electrodynamics.}
\label{vertexfigure}
\end{figure}
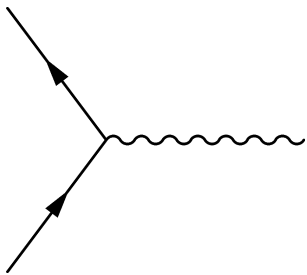

The above Hamiltonian gives a starting point from which one can begin doing quantum field theoretic calculations of state evolution (for example, in scattering problems).  This is, of course, a very complicated procedure.  For details as to how the Feynman rules for quantum electrodynamics can be derived from the above Hamiltonian, see \cite[sec.\ 17.9]{bjorkendrellfields}; \cite[ch.\ 8]{hatfield}; \cite[ch.\ 8]{weinbergQFT}; \cite[sec.\ 6.4]{tong}.  Note that when this is done the potential energy term in \eqref{QEDhamiltonian} is incorporated into the radiative vertex describing interactions (figure \ref{vertexfigure}).\footnote{To generate the standard radiative interaction vertex, the last term in \eqref{QEDhamiltonian} must be combined with part of the $\frac{\widehat{E}^2}{8 \pi}$ term.  Tong \cite[sec.\ 6.4.1]{tong} gives an alternative set of Feynman rules for quantum electrodynamics where the last term in \eqref{QEDhamiltonian} alone generates a non-standard radiative interaction vertex (and where there is also another kind of interaction).}  Depending on its orientation, this vertex can represent: (i) photon emission or absorption by an electron, (ii) electron-positron annihilation, (iii) electron-positron pair production, or (iv) photon emission or absorption by a positron.

\section{Conclusion}

In electrostatics and Newtonian gravity, we may use either potential energy or field energy to balance the work done on matter by the field and ensure conservation of energy.  In classical electrodynamics, we use field energy for this purpose.  Still, it is possible that, in addition to the energy in the electromagnetic field, there is also electromagnetic potential energy possessed by matter.  If we take the matter to be a classical Dirac field, then matter does have potential energy (as we saw by examining the symmetric energy-momentum tensor of the Dirac field).  This potential energy appears in the Hamiltonian of quantum electrodynamics, where it describes interactions between electrons, positrons, and photons.  Thus, the lesson of quantum electrodynamics seems to be that potential energies are indeed part of fundamental physics.

\vspace*{12 pt}
\noindent
\textbf{Acknowledgments}
Thank you to Jacob Barandes, Sean Carroll, Davison Soper, and, especially, Logan McCarty and the anonymous reviewers for helpful feedback and discussion.

\appendix

\section{Comparing Candidate Potential Energy Densities}\label{KPED}

In the Dirac field's energy density \eqref{diracenergydensity}, should we take (B) or (A)+(B) to be the potential energy density of the Dirac field?  Given the clarification in section \ref{potsec} that we are understanding potential energy to be energy of one entity that is dependent on the state of some other entity (in this case, energy of the Dirac field that is dependent on the state of the electromagnetic field), one might first think only (B) should be classified as potential energy density.  However, (B)'s dependence on the state of the electromagnetic field cannot be separated out from (A)+(B)'s dependence in a gauge-invariant way.  So, one could argue that (A)+(B) should be regarded as the potential energy density.

\begin{figure}[htb]
\center{\includegraphics[width=10 cm]{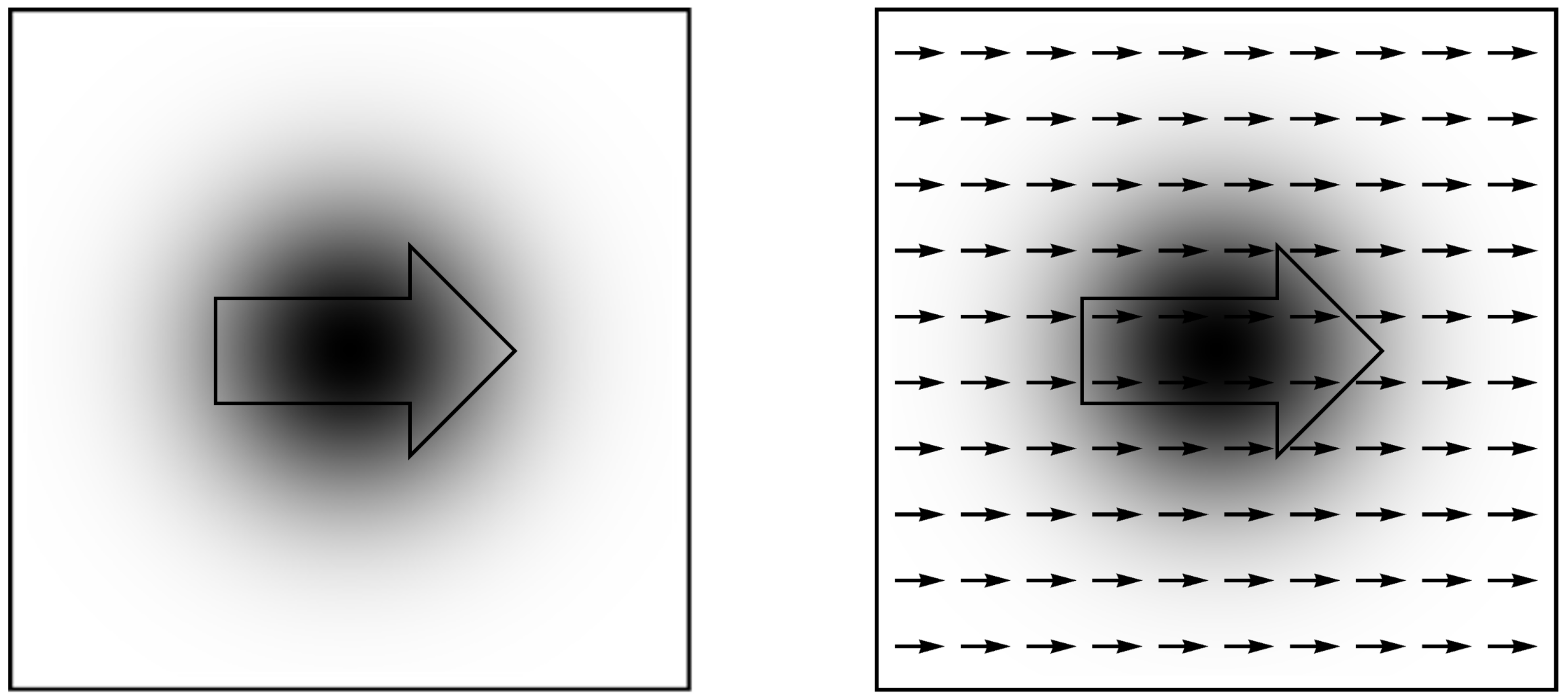}}
\caption{The first image shows the charge density of a wave packet in the Dirac field moving to the right in free space through a vector potential that is everywhere zero.  The second image shows the same situation in a different gauge, with arrows depicting a constant vector potential pointing to the right at every point in space.}
\label{Pfigure}
\end{figure}

To illustrate the difference between the two options, let us suppose that we have a Dirac field wave packet traveling forward in the $x$ direction in empty space (where there is no external electromagnetic field).  If we set aside the electromagnetic field created by the wave packet itself, then we can set the vector and scalar potentials to zero everywhere and have the Dirac field's potential energy density (B) vanish everywhere.  However, if we perform a gauge transformation \eqref{gaugetransformations} with $\alpha$ equal to some positive constant $C$ times $x$, the vector potential becomes $\vec{A} = (C,0,0)$ (depicted in figure \ref{Pfigure}).  This gauge transformation will change the Dirac field by an $x$-dependent phase factor of $e^{- i \frac{e C}{\hbar c} x}$, which leaves the charge and current densities unaltered.  The gauge transformation lowers (A) and increases (B), leaving (A)+(B) unchanged at every point.  To see that (B) is positive, note that the wave packet carries negative charge and is moving forward in the $x$ direction, so the current density $\vec{J}$ points in the negative $x$ direction (opposite the vector potential).  If (B) is the potential energy, then the wave packet has no potential energy in the first gauge and positive potential energy in the second gauge.  The potential energy is gauge-dependent, just as it was in Newtonian gravity and electrostatics.  On the other hand, if (A)+(B) is the potential energy then the packet (oddly) already possessed potential energy in the first gauge (where the potentials were zero everywhere) and carries the same potential energy in the second gauge.

Moving to a more complex case, let us consider the Aharonov-Bohm effect (in the context of classical field theory, not quantum physics).  Wallace \cite{wallace2014} has recognized the importance of the fact that gauge transformations affect both the matter field and the electromagnetic vector potential in the Aharonov-Bohm setup (though his focus is on a charged scalar field, not the Dirac spinor field).  From his analysis of the Aharonov-Bohm effect, Wallace concludes that the matter field and the vector potential are  ``interlinked,'' jointly representing the physical state of a single entity.  Although I have not collapsed the Dirac and electromagnetic fields into a single entity, I agree that, in an important way, the two fields are interlinked: the Dirac field has a potential energy density that depends on the physical state of the electromagnetic field.  But, is that potential energy density given by (B) or (A)+(B)?

\begin{figure}[htb]
\center{\includegraphics[width=8 cm]{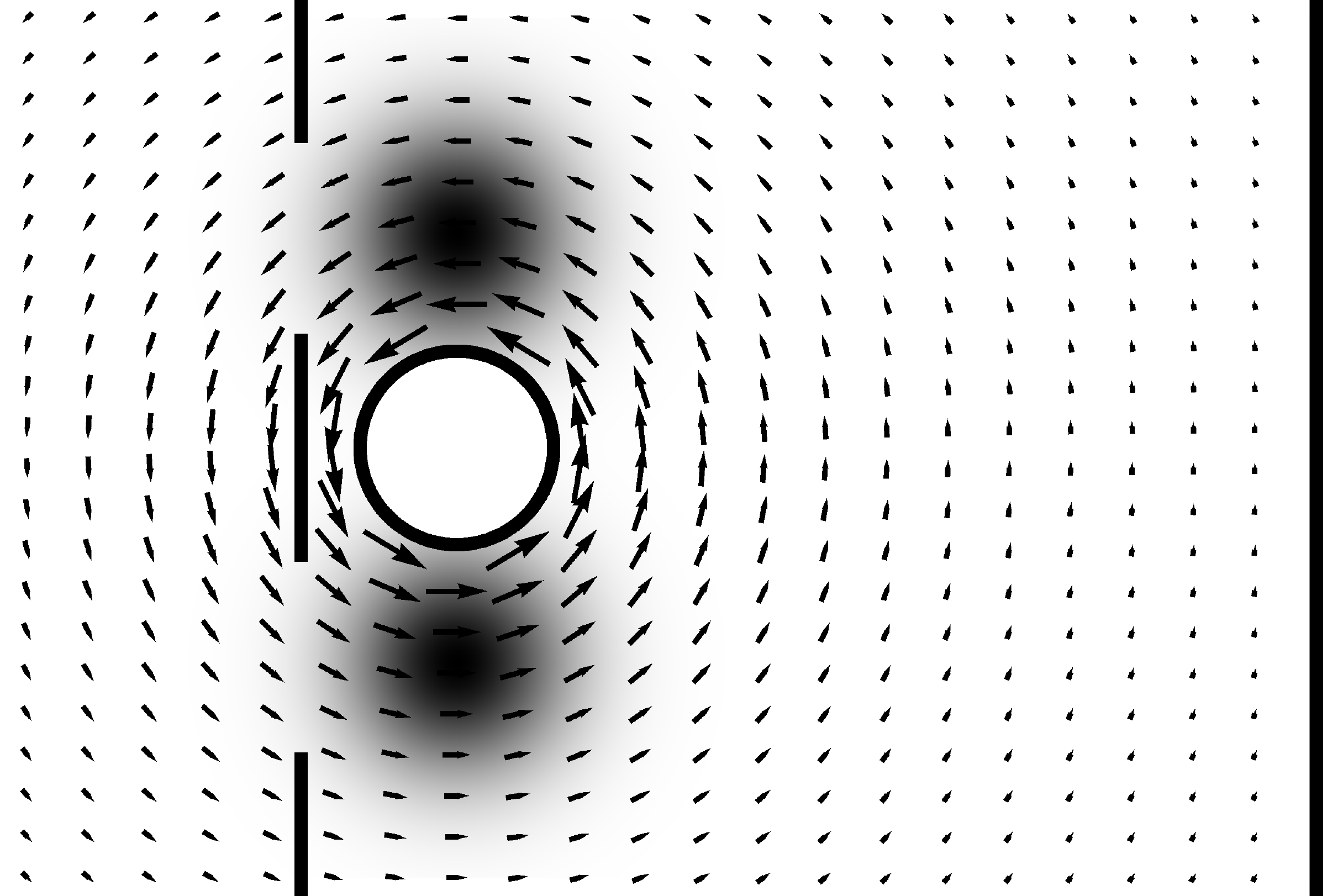}}
\caption{This image depicts charge densities for two wave packets that are traveling above and below a long solenoid (shown as a white circle), through a vector potential.  As time goes on, the wave packets will move toward the detection screen on the right while spreading and interfering.}
\label{ABfigure}
\end{figure}

The Aharonov-Bohm setup is illustrated in figure \ref{ABfigure}.  Let us suppose that the Dirac field is in a (classical) superposition of a wave packet that passes above the solenoid and a wave packet that passes below the solenoid.  Assuming the solenoid is long and carefully shielded, these wave packets pass through regions where the electromagnetic field is essentially zero.  However the vector potential is not negligible.  In the most natural choice of gauge, the vector potential circles the solenoid as depicted in figure \ref{ABfigure}.\footnote{For a mathematical description of this vector potential and the Aharonov-Bohm effect, see \cite[sec.\ 3.4]{ryder1996}.}  When the wave packets hit the screen, the interference pattern\footnote{In this classical field theory context, the interference pattern will be visible in the charge density of the Dirac field.} will depend on the strength of the magnetic field in the solenoid.  Seeking to avoid non-local interaction between the solenoid and the matter field, this shift in the interference pattern (observed in quantum experiments) has been taken as evidence that the scalar and vector potentials are more fundamental than electric and magnetic fields.  I will not attempt to sort out issues of locality here, or questions about the nature of the electromagnetic field.\footnote{See \cite{belot1998, lange, healey2007, wallace2014, maudlin2018} and references therein.}  Instead, I want to use this experiment to better understand our options for isolating the potential energy density of the Dirac field.

If the potential energy density is (B) and we adopt the choice of gauge in figure \ref{ABfigure}, the wave packet that passes underneath the solenoid will acquire a positive potential energy because its current points opposite the vector potential.  Accompanying that potential energy will be a shift in the phase of the wave packet from what it would have been had the packet been moving at the same velocity in the absence of a vector potential (just as in the case from figure \ref{Pfigure}).  The wave packet that passes above the solenoid will acquire a negative potential energy and an opposite shift in the phase.  These phase shifts are responsible for the interference pattern.  In a different gauge, the potential energies of the wave packets would be different and so would the phase shifts.  But, the interference pattern would be the same.  Thus, taking (B) to be the potential energy density, we can attribute the shift in the interference pattern to a difference in potential energies between the wave packets.  Although we can push that bump in the rug around through gauge transformations, it cannot be eliminated.  By contrast, if we take (A)+(B) to be the potential energy density, then the potential energies of the two wave packets would be just the same as if they were propagating in free space.  The predicted interference pattern is the same, but it cannot be understood as the result of the different wave packets having different potential energies.

I do not have a definitive argument to offer in favor of taking (B) to be the potential energy density, but it is the option that I lean towards and thus I have chosen to treat (B) as the potential energy density of the Dirac field in the main text of the article.  Taking (B) to be the potential energy density gives an attractive understanding of the two cases described above.  This potential energy density is gauge-dependent, but that feature is shared with the potential energy densities used in Newtonian gravity and electrostatics (see section \ref{ESsection}).

\end{document}